\documentclass[runningheads]{llncs}
\usepackage[T1]{fontenc}
\usepackage{graphicx}
\usepackage{subcaption}
\usepackage{float}
\usepackage{svg}

\begin{document}
\title{Lightweight Target-Speaker-Based Overlap Transcription for Practical Streaming ASR}
\titlerunning{Lightweight Target-Speaker-Based Overlap Transcription}
\author{Ale\v{s} Pra\v{z}\'ak\orcidID{0000-0001-9453-0034} \and
Marie Kune\v{s}ov\'a\orcidID{0000-0002-7187-8481} \and
Josef Psutka\orcidID{0000-0002-0764-3207}}

\authorrunning{A. Pra\v{z}\'ak et al.}

\institute{University of West Bohemia in Pilsen, Pilsen, Czech Republic\\
\url{https://www.zcu.cz/en}}

\maketitle

\begin{abstract}
Overlapping speech remains a major challenge for automatic speech recognition (ASR) in real-world applications, particularly in broadcast media with dynamic, multi-speaker interactions. We propose a light\-weight, target-speaker-based extension to an existing streaming ASR system to enable practical transcription of overlapping speech with minimal computational overhead. Our approach combines a speaker-independent (SI) model for standard operation with a speaker-conditioned (SC) model selectively applied in overlapping scenarios. Overlap detection is achieved using a compact binary classifier trained on frozen SI model output, offering accurate segmentation at negligible cost. The SC model employs Feature-wise Linear Modulation (FiLM) to incorporate speaker embeddings and is trained on synthetically mixed data to transcribe only the target speaker. Our method supports dynamic speaker tracking and reuses existing modules with minimal modifications. Evaluated on a challenging set of Czech television debates with 16\% overlap, the system reduced WER on overlapping segments from 68.0\% (baseline) to 35.78\% while increasing total computational load by only 44\%. The proposed system offers an effective and scalable solution for overlap transcription in continuous ASR services.

\keywords{Streaming ASR \and Overlapping Speech \and Speaker Conditioning.}
\end{abstract}

\section{Introduction}

Today, automatic speech transcription reaches very high accuracy even for minority languages. However, transcription errors persist, even under ideal acoustic conditions, such as in the case of television and radio broadcasting. According to our expertise, there are two primary sources of errors in this case – uncommon words (such as foreign proper or geographical names) and overlapping speech. While uncommon words introduce mainly isolated misrecognitions, particularly in news content, overlapping speech may easily compromise the intelligibility of whole sentences, especially during passionate debates. In the ETAPE Evaluation Campaign (French Broadcast News and Debates)~\cite{6639163}, the overlapping speech constituted approximately 5.9\% of the total speech content, peaking at almost 9\% in the debates. This may be a considerable source of transcription errors for conventional systems, which can transcribe only the dominant speaker at best. Given that our partner company operates a continuous 24/7 streaming transcription service across more than 50 TV and radio channels in the Czech and Slovak languages, we sought a lightweight and practical enhancement to support overlap transcription and further reduce our mean word error rate (WER), which currently stands slightly above 3\%.

Prior work on automatic transcription of overlapping speech has explored several paradigms with varying levels of suitability for streaming practical systems. Speech separation approaches, such as~\cite{10094612}, aim to isolate individual speech signals before transcription, offering strong performance, but often requiring significant computational resources and introducing latency challenges. In streaming contexts, the number of speakers is usually not known a priori, and assigning speech to the correct source (speaker tracking) is difficult due to permutation ambiguity~\cite{7952154}. Target speaker extraction (TSE) methods~\cite{elminshawi2023newinsightstargetspeaker} leverage prior speaker information to extract relevant speech, reducing complexity compared to complete separation. TSE models often rely on bidirectional context, e.g., BiLSTMs, which is incompatible with streaming~\cite{Zmolikova_2023} and can be computationally heavy, especially with large input windows and complex architectures. In addition, the TSE output may not fully conform to the acoustic model, which is typically trained on the original acoustics. More recently, end-to-end multispeaker ASR models, often using serialized output training (SOT)~\cite{kanda2022streamingmultitalkerasrtokenlevel}, directly transcribe overlapping speech without intermediate separation, enabling tighter integration and lower latency, though often at the cost of performance or generalizability in real-time applications. Although these approaches have advanced the field, practical deployment in streaming systems remains constrained by trade-offs between accuracy, latency, and system complexity.

Target speaker transcription, usually referred to as Target Speaker Automatic Speech Recognition (TS-ASR)~\cite{10095115}, offers a promising solution for streaming transcription systems where the number of speakers is not known in advance, but their characteristics can be obtained on the fly. We adopted some basic ideas from~\cite{10097139} and enhanced our practical streaming ASR system to deal with overlapping speech. Our proposed approach can be integrated into standard ASR pipelines with low effort and computational overhead.

\section{Baseline system}

We first describe our baseline system, as it is built entirely in-house without relying on any publicly available ASR systems or standardized datasets.

We have access to a large corpus of manually transcribed broadcasting comprising more than 10.000 hours. However, these transcriptions were originally created for editorial purposes, not for machine learning. As a result, they are not verbatim and include typographical errors and irregularities such as reordered words. To prepare the data, we performed automatic alignment of the audio with the non-verbatim transcriptions and selected only the longest continuous segments (across speakers) with high confidence. The resulting train dataset contains over 6.500 hours of aligned data, including approximately 1.000 hours of non-speech segments (e.g., music, jingles). Each transcription is also associated with speaker identity information, though these annotations are noisy. Therefore, we applied cross-verification across speaker turns to remove unreliable samples, resulting in a speaker-labeled corpus featuring over 20,000 distinct speakers with varying amounts of data.

A practical streaming ASR system typically consists of a transformer-based acoustic model (e.g., wav2vec 2.0~\cite{baevski2020wav2vec20frameworkselfsupervised}, Conformer~\cite{gulati2020conformerconvolutionaugmentedtransformerspeech}, or Zipformer~\cite{yao2024zipformerfasterbetterencoder}), a decoder optionally supported by a language model, streaming speaker diarization (including speaker change detection and speaker identification), and finally a punctuation restoration module. In our system, we use a base-sized wav2vec 2.0 model trained for continuous recognition, jointly optimized for speaker change detection. Since the system operates on uninterrupted audio streams, we segment the audio into 15-second windows with 3-second overlaps on both sides. Training mirrors this setup: long training segment (up to 360 seconds, based on GPU capacity) is unfolded with overlaps to form a mini-batch, the forward pass is computed, and resulting logits are aggregated from 9-second operation windows before being passed to the CTC loss function using the reference transcription of the full segment. This approach minimizes the need for forced inaccurate segmentation and preserves contextual information. The model was fine-tuned using the Czech monolingual pre-trained model ClTRUS~\cite{lehecka22_interspeech}.

To support online speaker change detection with no additional delay and negligible computational overhead, we introduce an additional per-frame binary classification head in the model. This is trained using the collar-aware binary cross-entropy loss as proposed in~\cite{kalda2022collarawaretrainingstreamingspeaker}. Unlike the authors, we do not use a fixed-size collar around annotated speaker changes. Instead, a speaker change label may appear anywhere between the last word of one speaker and the first word of the next, based on timestamps from the alignment process. To account for the limited input window of the model, we label non-speech segments longer than the context length (3 seconds) as separate speaker turns at both boundaries. This allows for effective separation of music or jingles. If the same speaker appears on both sides of the non-speech segment, their turns can be unified later through speaker identification. On real broadcast data, our speaker change detection achieves an F1 score greater than 85\%.

For speaker identification, we extract speaker embeddings using an NVIDIA TitaNet-S model~\cite{9746806}, which we re-trained on our train dataset. To address speaker imbalance, data from frequent speakers were limited. During system operation, speaker identification is performed iteratively, beginning with an initial 2-second audio snippet. Although we have a large set of known speaker identities, new speakers are continuously added if none of the stored embeddings sufficiently match the current one.

Our ASR decoder is a proprietary, optimized CTC decoder that incorporates a 4-gram language model with a vocabulary exceeding 1.5 million words and support for unknown word transcription. Finally, punctuation restoration is applied to structure the recognized text into sentences using full stops and question marks, and to add commas, which are especially prevalent in Slavic languages. This module is based on our RoBERTa model with a subword vocabulary of 100,000 tokens.

The entire system is implemented in C\texttt{++} and can operate on both CPU and GPU, depending on hardware availability, to ensure efficient large-scale deployment.

\section{Multi-speaker scenario}

\begin{figure}
  \centering
  \resizebox{\columnwidth}{!}{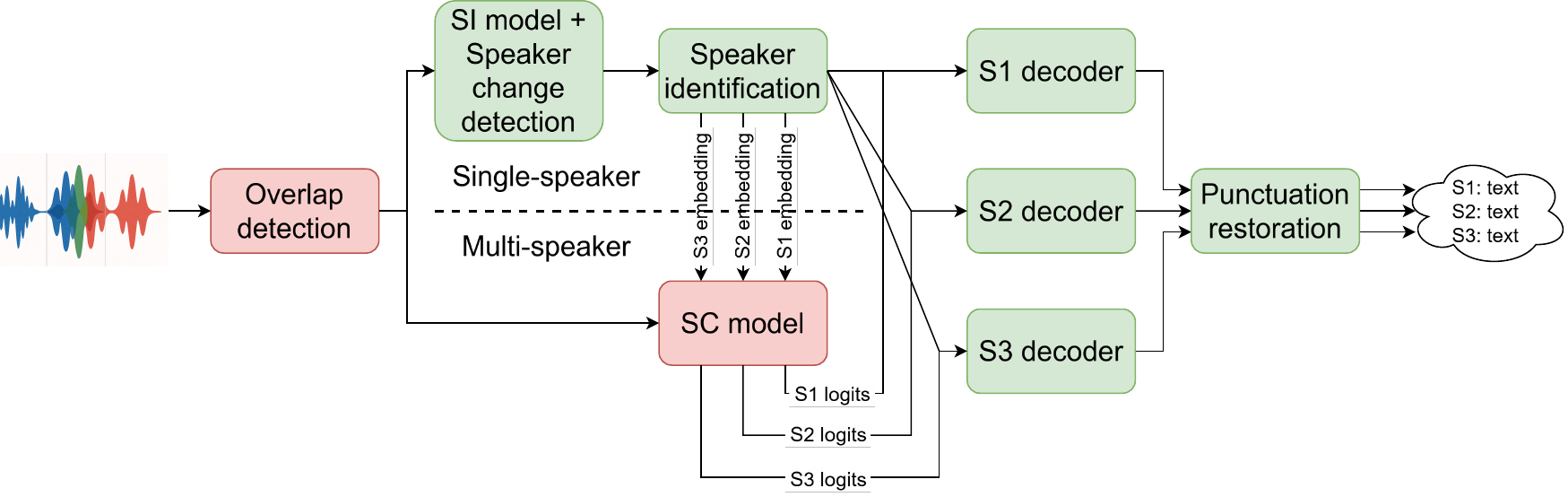}
  \caption{Our system schema with already existing modules in green and proposed modules for the multi-speaker scenario in red.}
  \label{figure1}
\end{figure}

Our goal was to extend the existing streaming ASR system to handle multi-speaker segments, to some extent, with minimal computational overhead and without degrading performance on single-speaker speech. Given that overlapping speech occurs only during a small fraction of the total audio, our strategy is to use an existing single-speaker acoustic model, hereinafter referred to as Speaker-Independent (SI) model, most of the time, and in case a multi-speaker scenario is detected, an acoustic model conditioned by a target-speaker embedding - a Speaker-Conditioned (SC) model - is used more than once. Importantly, the rest of the system remains largely unchanged. An overview of the proposed architecture is shown in Fig.~\ref{figure1}.

An overlap detection module determines whether the system operates in a single-speaker or multi-speaker mode. In the single-speaker case (i.e., only one speaker is active at any given time, though speakers may change), the current audio window is processed by the SI model. Speaker change boundaries are detected using the model’s speaker change detection head, and the resulting segments are routed to separate decoders based on speaker identification, each segment being handled by a single decoder.

When overlapping speech is detected, the Speaker-Conditioned (SC) model is applied multiple times to the same input window, each instance conditioned on a different speaker embedding previously extracted by the speaker identification module. The resulting logits are then fed in parallel to the corresponding speaker-attributed decoders. The number of co-existing decoders limits the maximum number of simultaneous speakers the system allows.

Although the decoders are speaker-attributed, they dynamically switch affiliations based on speaker activity. Most of the time, during single-speaker segments, only one decoder remains active, and the decoding context is seamlessly transferred between decoders at speaker change points. Finally, individual speaker streams are processed by the punctuation restoration module and merged into the final transcript. For our application, sentence-level interleaving is generally sufficient.

\subsection{Overlap detection}

The overlap detection module is a critical component of our system, enabling dynamic switching between single-speaker and multi-speaker processing modes. Numerous approaches have been proposed for overlap detection, including the use of wav2vec 2.0-based models~\cite{kunesova2024}. The authors claim that wav2vec 2.0 models, particularly when fine-tuned on in-domain data, exhibit strong performance in speaker overlap detection, often outperforming traditional methods.

Despite having access to a large amount of in-domain data, we lack sufficient real-world multi-speaker data with overlap annotations for training. Consequently, we trained on artificially mixed data. We used the same dataset as for our baseline system, but with additional segmentation at speaker boundaries to ensure that each segment contained only a single speaker. The maximum segment length was limited to 45 seconds. During training, two randomly selected segments were mixed with random temporal shift and volume scaling to simulate overlapping speech. Additional details are provided in the next section.

Initially, we trained a tiny wav2vec 2.0 model with a per-frame binary classifier from scratch using artificially mixed data. However, its performance in real overlapping speech was inadequate for practical implementation. The model predictions were overly sensitive, resulting in numerous short-segment false positives and false negatives. To address this, we adopted a simpler yet surprisingly effective approach. We reused our baseline SI acoustic model and appended a binary classification head for overlap detection, while keeping the rest of the model (including all transformer layers) frozen. As a result, only 769 parameters were trained. This minimal head learned to distinguish overlapping speech based solely on the uncertainty of the model’s output, as expressed through its logit distribution. In single-speaker scenarios, where the SI model achieves a very low word error rate, its output is highly confident. In contrast, overlapping speech (unseen during SI training) produces a markedly different and less confident logit distribution, which the classification head effectively exploits. The resulting model yields compact segmentation and successfully detects even short overlap events.

The final decision of the overlap detection module incorporates two post-processing steps to improve robustness:

\begin{itemize}
\item Segments labeled as single-speaker but shorter than one second are relabeled as overlapping, as speaker identity cannot be reliably determined over such short durations.
\item Segments labeled as overlapping but shorter than one second and flanked by the same speaker on both sides are relabeled as single-speaker. These typically represent brief interjections (e.g., one-word overlaps) that a speaker-conditioned model is not expected to transcribe effectively.
\end{itemize}

A major advantage of this method is its negligible computational overhead, making it suitable for online streaming applications.

\subsection{Speaker-conditioned model}

To enable seamless switching between the SI and SC models while maintaining compatibility of the output logits, we used the same wav2vec 2.0 architecture for the SC model. To condition the model on a target speaker, we applied Feature-wise Linear Modulation (FiLM)~\cite{FiLM} to incorporate the embedding of the target speaker. Specifically, FiLM was applied in the first transformer block following the feature extractor. The model is trained to transcribe only the utterances of the target speaker if present in a speech mixture and to produce an empty transcription if the target speaker is absent. This makes training particularly challenging, especially as we did not start from a pre-trained model.

During training, we synthesized overlapping speech by mixing two randomly selected segments from different speakers with a random temporal offset. Given that segment lengths ranged from 1 to 45 seconds, this setup covered two typical overlap scenarios: speaker transitions with a variable-length overlap, and interruptions, where one speaker interjects briefly, but the first continues speaking. In the second scenario, the volume of the inserted utterance was randomly adjusted to simulate cases such as shouting or subdued self-talk. 

Speaker embeddings for the overlapping speakers were selected in various ways. For each speaker, we experimented with using: a random embedding from their available segments, the mean embedding from several random segments, and the embedding from the immediately preceding segment. The best performance was achieved using the embedding of the immediately preceding segment, likely due to temporal variability in speaker characteristics across our dataset, which spans more than a decade.

Although this approach enables transcription of overlapping speech, in practice, the identities of the overlapping speakers are unknown. Therefore, during training, we simulate realistic inference conditions by conditioning the model on embeddings of all participating speakers, as well as an additional randomly selected "non-speaking" speaker. As in the baseline model training, the mixed audio signal is unfolded into overlapping windows to create a mini-batch. The SC model is then applied three times per mini-batch (once per speaker embedding) and the CTC loss is computed for each run.

To address the substantial loss imbalance between speaking and non-speaking cases, we normalize the CTC loss for the non-speaking speaker by the number of inserted characters (i.e., frames for which the argmax over logits is any speech label). This normalization factor naturally decreases as training progresses. The final loss for the mini-batch is the sum of the normalized losses across all speakers.

\section{Test data}

Our test dataset comprises ten one-hour television debate episodes, each featuring one host and six or more guests. These episodes were intentionally selected for their high occurrence of passionate discussion and frequent speech overlaps. Independent speaker transcriptions were manually created using the ELAN annotation tool~\cite{ELAN}.

\begin{figure}
  \centering
  \begin{subfigure}[b]{0.33\textwidth}
    \includegraphics[width=\textwidth]{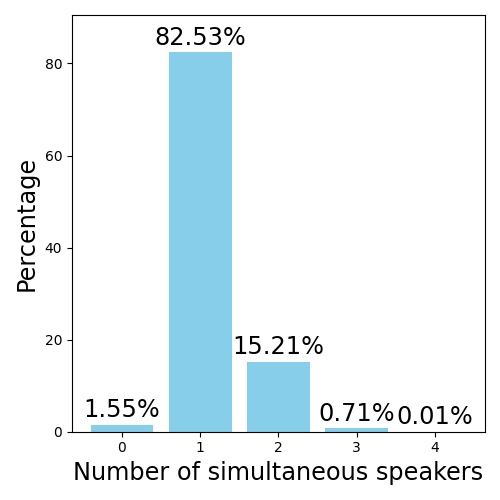}
    \caption{Distribution of the number of simultaneous speakers in test data.}
    \label{figure2a}
  \end{subfigure}
  \hfill
  \begin{subfigure}[b]{0.66\textwidth}
    \includegraphics[width=\textwidth]{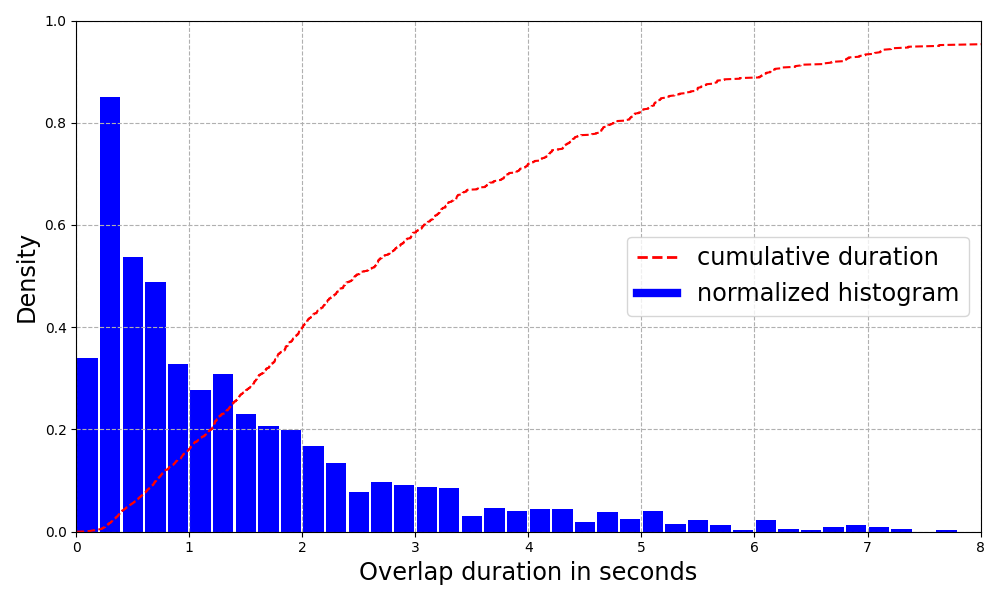}
    \caption{Histogram of overlap durations.}
    \label{figure2b}
  \end{subfigure}
  \caption{Distribution of overlap durations in test data.}
\end{figure}

Speech overlaps account for almost 16\% of the total duration of the test set (see Fig.~\ref{figure2a}), representing a deliberately challenging scenario compared to typical broadcast content. The distribution of overlap durations is presented in Fig.~\ref{figure2b}. The cumulative duration of overlaps shows that overlaps longer than one second cover about 84\% of the cumulative duration of overlapping speech, although they constitute only 49\% of all overlap instances.

The test data corpus will be made available free of charge for non-commercial use via the LINDAT/CLARIAH-CZ repository, pending resolution of the copyright issues related to the audio recordings.

\section{Experiments}

Baseline results were obtained under a single-speaker scenario, assuming that the overlap detection module consistently produced negative results, and using a single decoder. As reference transcriptions are speaker-specific, words spoken during overlapping segments were manually attributed to individual speakers for evaluation, since our standard speaker identification module does not support overlapping speech. These results are reported in the row labeled "SI model (BASELINE)" in Table~\ref{tab1}. The overall WER was 19.80\%, with 3.74\% WER on single-speaker segments and 68.00\% on overlapping segments.

In multi-speaker scenarios, the SC model is applied multiple times to the same audio window, each time conditioned on a different target-speaker embedding. Since new speakers appear frequently in a streaming context (often contributing only a sentence or two), it is necessary to limit the number of speaker embeddings processed by the SC model. A natural strategy is to use the $N$ most recent speakers. As $N$ increases, more overlaps can be transcribed correctly, although at the cost of increased computational complexity.

Because the performance of the SC model is strongly influenced by the quality of the target-speaker embedding, we evaluated four methods for selecting it:

\begin{itemize}
\item MEAN - mean embedding from all previous utterances by the target speaker.
\item MEDIAN - component-wise median of embeddings from previous utterances by the target speaker.
\item MEDOID - medoid of the speaker’s embeddings, i.e., the embedding with the smallest average distance to the others.
\item LAST - embedding obtained from the speaker’s most recent utterance.
\end{itemize}

We conducted a series of experiments varying both the number of the most recent speakers ($N$) and the embedding selection method. The number of decoders was set to $N$ as well. Table~\ref{tab1} summarizes the resulting WERs and the relative computational complexity in terms of CPU/GPU time (HW), normalized to 1 for the baseline system.

\renewcommand{\arraystretch}{1.25}
\begin{table}[H]
\centering
\caption{Word Error Rate (WER, \%) and relative computational complexity (HW) for different values of $N$ (the number of most recent speakers considered) and various methods of target-speaker embedding selection.}
\label{tab1}
\begin{tabular}{l@{\hspace{10pt}}c@{\hspace{10pt}}c@{\hspace{10pt}}c@{\hspace{10pt}}c@{\hspace{10pt}}c@{\hspace{10pt}}c}
\hline
\textbf{Model} & \textbf{N} & \textbf{HW} & \multicolumn{4}{c}{\textbf{WER}} \\ \hline \hline
SI model (BASELINE) & 1 &   1  & \multicolumn{4}{c}{19.80}     \\ \hline
                    &   &      & MEAN  & MEDIAN & MEDOID & LAST \\ \cline{4-7}
                    & 2 & 1.29 & 14.56 & 14.37  & 14.34  & 15.04 \\
SI + SC models      & 3 & 1.40 & 12.30 & 12.45  & 11.84  & 13.27 \\
                    & 4 & 1.44 & 12.36 & 12.51  & \textbf{11.75} & 13.24 \\ \hline
SC model only       & 4 & 1.44 & 12.97 & 13.18  & 13.73  & 15.41 \\ \hline
\end{tabular}
\end{table}

The MEDOID method for target-speaker embedding selection achieved slightly better performance than the other approaches. This may be attributed to its robustness against outlier embeddings, which can result from speaker misidentifications or very short utterances.

Both wav2vec 2.0 models (the SI model with speaker change and overlap detection heads, and the SC model) share the same feature encoder, which is typically not fine-tuned. As a result, the associated CNN layers are computed only once per signal window, regardless of how many forward passes are required. Increasing $N$ beyond three yields minimal gains in WER, also because overlaps involving four simultaneous speakers are extremely rare in our data, but it still adds computational overhead.

To assess the benefit of dynamically switching between single-speaker and multi-speaker scenarios, we also evaluated a configuration labeled 'SC model only', where the SI model was replaced by the SC on single-speaker segments, using the embedding of the corresponding speaker. This comparison highlights the advantage of using the SI model for the majority of the audio while relying on the SC model only when an overlap is detected.

\section{Conclusion}

We proposed a practical streaming transcription architecture capable of handling overlapping speech, to some extent, with minimal computational overhead. Our approach enhances a standard transcription system with two additional modules: an overlap detection module and a speaker-conditioned acoustic model. Both components can be efficiently trained and integrated into existing pipelines.

We achieved a significant reduction in overall WER on our test dataset, lowering it from 19.80\% to 11.75\%. Importantly, the transcription accuracy for single-speaker segments remains unaffected, while the WER for overlapping segments improved from a baseline of 68.00\% to 35.78\% in our best-case experiment. In addition to this performance gain, our proposed system is capable of attributing recognized words in multi-speaker segments to the correct speakers - an ability the baseline system lacked. These results demonstrate the viability of our system in real-world streaming applications where occasional overlapping speech occurs, without sacrificing performance or efficiency in more typical single-speaker scenarios.

\begin{credits}

\subsubsection{\ackname} This work was supported by the project\break CZ.02.01.01/00/23\_021/0008436, co-financed by the European Union and funded by the Ministry of Education, Youth and Sports of the Czech Republic through the Operational Programme Johannes Amos Comenius (OP JAK).

\end{credits}

\bibliographystyle{splncs04}
\bibliography{references}

\end{document}